\documentstyle[12pt,aaspp4,lscape,overcite,psfig,flushrt]{article}

\renewcommand{\thebibliography}[1]{\clearpage\subsection*{REFERENCES}\list
 {\arabic{enumi}.}{\settowidth\labelwidth{[#1]}\leftmargin\labelwidth
 \advance\leftmargin\labelsep
 \usecounter{enumi}}
 \def\newblock{\hskip .11em plus .33em minus .07em}
 \sloppy\clubpenalty4000\widowpenalty4000
 \sfcode`\.=1000\relax}
 
\def\aanat#1{{\it Astron. Astrophys.\/} {\bf #1}}
\def\ajnat#1{{\it Astron. J.\/} {\bf #1}}
\def\apjnat#1{{\it Astrophys. J.\/} {\bf #1}}
\def\apjsupnat#1{{\it Astrophys. J. Suppl. Ser.\/} {\bf #1}}

\def\mnnat#1{{\it Mon. Not. R. Astron. Soc.\/} {\bf #1}}
\def\natnat#1{{\it Nature\/} {\bf #1}}
\def\pasjnat#1{{\it Publ. Astron. Soc. Jpn\/} {\bf #1}}
\def\prlnat#1{{\it Phys. Rev. Lett.\/} {\bf #1}}

\begin{document}

\title{Resolving the extragalactic hard X-ray background
}

\author{R. F. Mushotzky$^{\ast}$, L. L. Cowie$^{\dag}$,
A. J. Barger$^{\dag}$, K. A. Arnaud$^{\ast\ddag}$}

\leftline{$^{\ast}$\ NASA Goddard Space Flight Center, Code 662,
Greenbelt, MD 20771}

\leftline{$^{\dag}$\ Institute for Astronomy, University of Hawaii, 2680
Woodlawn Drive, Honolulu, HI 96822}

\leftline{$^{\ddag}$\ Astronomy Department, University of Maryland,
College Park, MD 20742}

\vskip 0.2cm

\centerline{To be published in {\it Nature}}


{\bf
The origin of the hard (${\bf 2-10}$\ keV) X-ray background has 
remained mysterious for over 35 years. Most of the soft
(${\bf 0.5-2}$\ keV) X-ray background has been resolved into 
discrete sources, which are primarily quasars; however, these 
sources do not have the flat spectral shape required to 
match the X-ray background spectrum. Here we report the results 
of an X-ray survey 30 times more sensitive than previous studies 
in the hard band and four times more sensitive in the soft
band. The sources detected in our survey account for at least 
75 per cent of the hard X-ray background. The mean X-ray 
spectrum of these sources is in good agreement with that of the 
background. The X-ray emission from the majority of the detected 
sources is unambiguously associated with either the nuclei
of otherwise normal bright galaxies or optically faint sources,
which could either be active nuclei of dust enshrouded 
galaxies or the first quasars at very high redshifts.
}

For some time after the discovery of the
cosmic X-Ray background (XRB)\cite{giacconi}, there was
considerable controversy over whether the background
arose from a superposition of discrete sources or from
thermal bremsstrahlung emission from a hot intergalactic gas.
We now know that the bulk of the XRB cannot originate in
a uniform hot intergalactic medium
since a strong Compton distortion on the cosmic microwave
background spectrum
was not observed by the FIRAS instrument on {\it COBE}
\cite{mather2,wright}.

At soft X-ray energies ($0.5-2$\ keV) the XRB has been extensively
studied with the {\it ROSAT} satellite. The deepest {\it ROSAT}
source counts reach $\sim 1000$\ per square degree at a limiting
flux of $10^{-15}$\ erg\ cm$^{-2}$\ s$^{-1}$, and at this level
$70-80$ per cent of the XRB is resolved into discrete sources
\cite{hasinger}. The great majority of the optical identifications
of a complete sample of 50 {\it ROSAT} sources, at
a limiting flux of $5\times 10^{-15}$\ erg\ cm$^{-2}$\ s$^{-1}$,
are unobscured active galactic nuclei (AGN) \cite{schmidt1}.
However, because the objects detected in the soft band do not have
the spectrum of the XRB, a new population of absorbed or
flat spectrum objects are needed to make up the background at
higher energies. Detailed models developed
to resolve this ``spectral paradox'' assumed that
most of the flux in the XRB is produced by active galaxies that 
are obscured by dust. When deep imaging sky surveys with the 
{\it ASCA}\cite{ueda,ueda1,ueda2,cagnoni} and {\it BeppoSAX}\cite{fiore} 
satellites became possible in the hard ($>2$\ keV) X-ray band,
$\sim 30$\ per cent of the hard XRB was resolved, but only 
indirect identifications of the optical counterparts could be made.

The {\it Chandra} satellite\cite{weisskopf}, 
with its great sensitivity over a wide energy range,
excellent image quality, superb positional accuracy,
and reasonable field-of-view, can directly image the sources that
make up the hard XRB. We have therefore carried out a deep imaging 
survey of the Hawaii Deep Survey Field SSA13 with the ACIS-S 
instrument on {\it Chandra} to resolve
the hard XRB and to identify the nature of the sources that produce it.
We chose to centre on the SSA13 field\cite{lilly}, which has existing
multiwavelength observations\cite{windhorst,songaila,barger},
to maximize the immediate identification of optical/near-infrared (NIR)
counterparts and redshifts for the X-ray source detections.
We find that above a flux threshold of
$2.5\times 10^{-15}$\ erg\ cm$^{-2}$\ s$^{-1}$ ($2-10$\ keV),
we can account for at least $75$ per cent of the sky flux,
with the main uncertainty being the sky flux itself.
Our deep optical observations show a rich assortment
of hard X-ray sources which could not have been discovered by
previous satellites.

\vskip 0.5cm
\centerline{\bf {\it Chandra} X-ray Survey of SSA13}
\vskip 0.2cm

The SSA13 observation was performed on 1999 December 3--4 for an
elapsed time of 100.9\ ks. The optical axis of the telescope at
RA(2000)$=13^h\ 12^m\ 21.40^s$,
Dec(2000)$=42^{\circ}\ 41^{'}\ 20.96^{''}$ was positioned on
the back illuminated CCD (S3) of ACIS since
this detector has a much better soft X-ray sensitivity than the
front illuminated chips. Furthermore, since the back illuminated
detectors did not suffer the radiation damage which affected the
front illuminated chips in orbit, they are well
characterised by extensive ground-based calibrations.

The overall sensitivity of the instrument spans
a wide energy range from 0.2 to 10\ keV.
Two energy-dependent images of the S3 chip were generated in
the hard ($2-10$\ keV) and soft ($0.5-2$\ keV) bands, as was
a $2-10$\ keV image of the front illuminated S2 chip that
covered a neighboring region.
We extracted sources independently for the hard
and soft band images. Sources brighter than 
$3.2\times 10^{-15}$\ erg\ cm$^{-2}$\ s$^{-1}$ ($2-10$\ keV) 
or $3\times 10^{-16}$\ erg\ cm$^{-2}$\ s$^{-1}$ ($0.5-2$\ keV; S3 chip
only) which lie within 6 arcminutes of the optical axis are 
given in Table~1, ordered by right ascension; the table contains 22 
sources selected in the hard band and a further 15 sources 
selected solely in the soft band.
Details of the extraction and calibration of the X-ray data and
of the optical photometry may be found in the table footnote.

\vskip 0.5cm
\centerline{\bf Number Counts and the Resolution of the X-ray Background}
\label{secbkgd}
\vskip 0.2cm

\begin{figure}
\centerline{\psfig{file=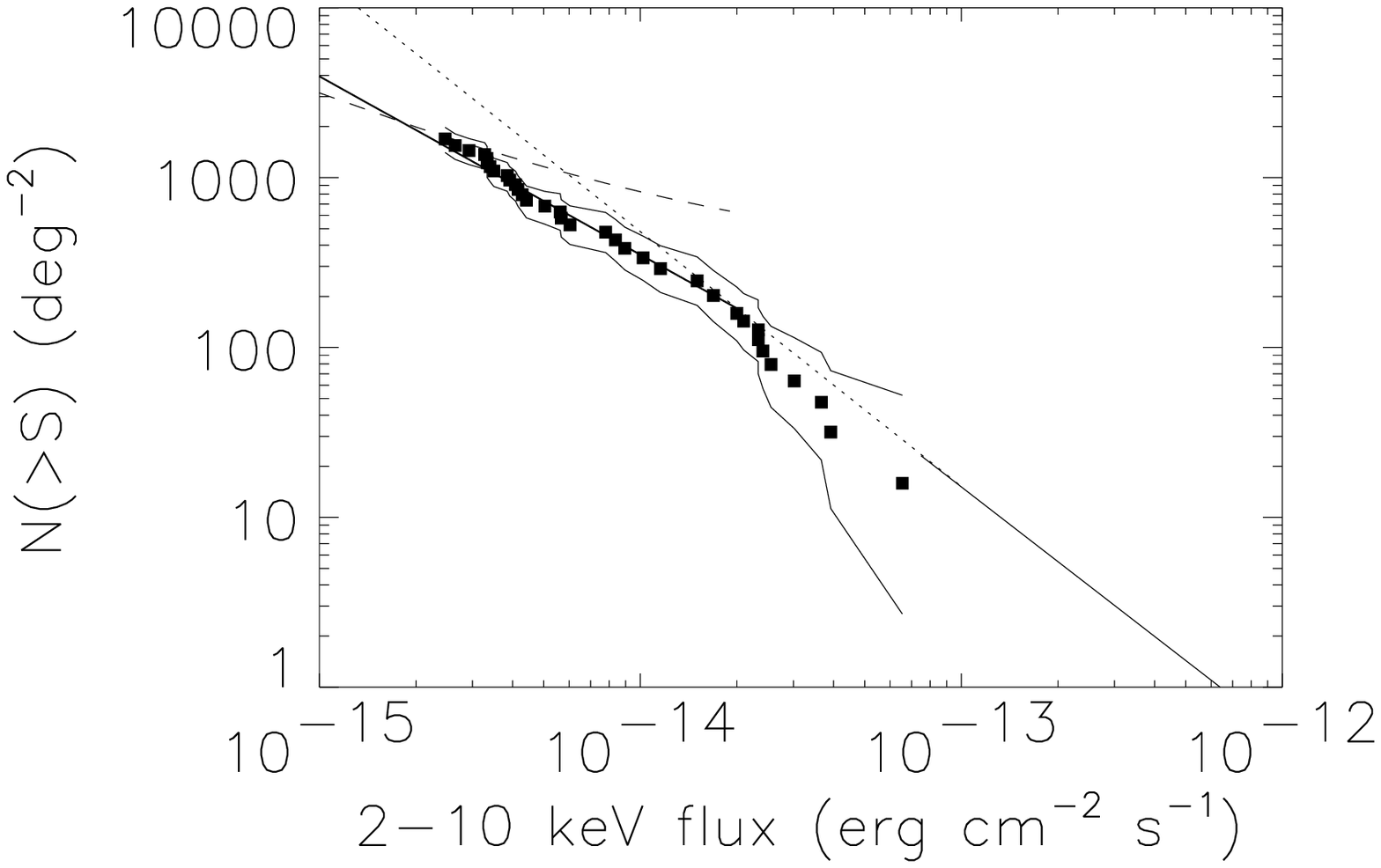,height=2.3in}
\psfig{file=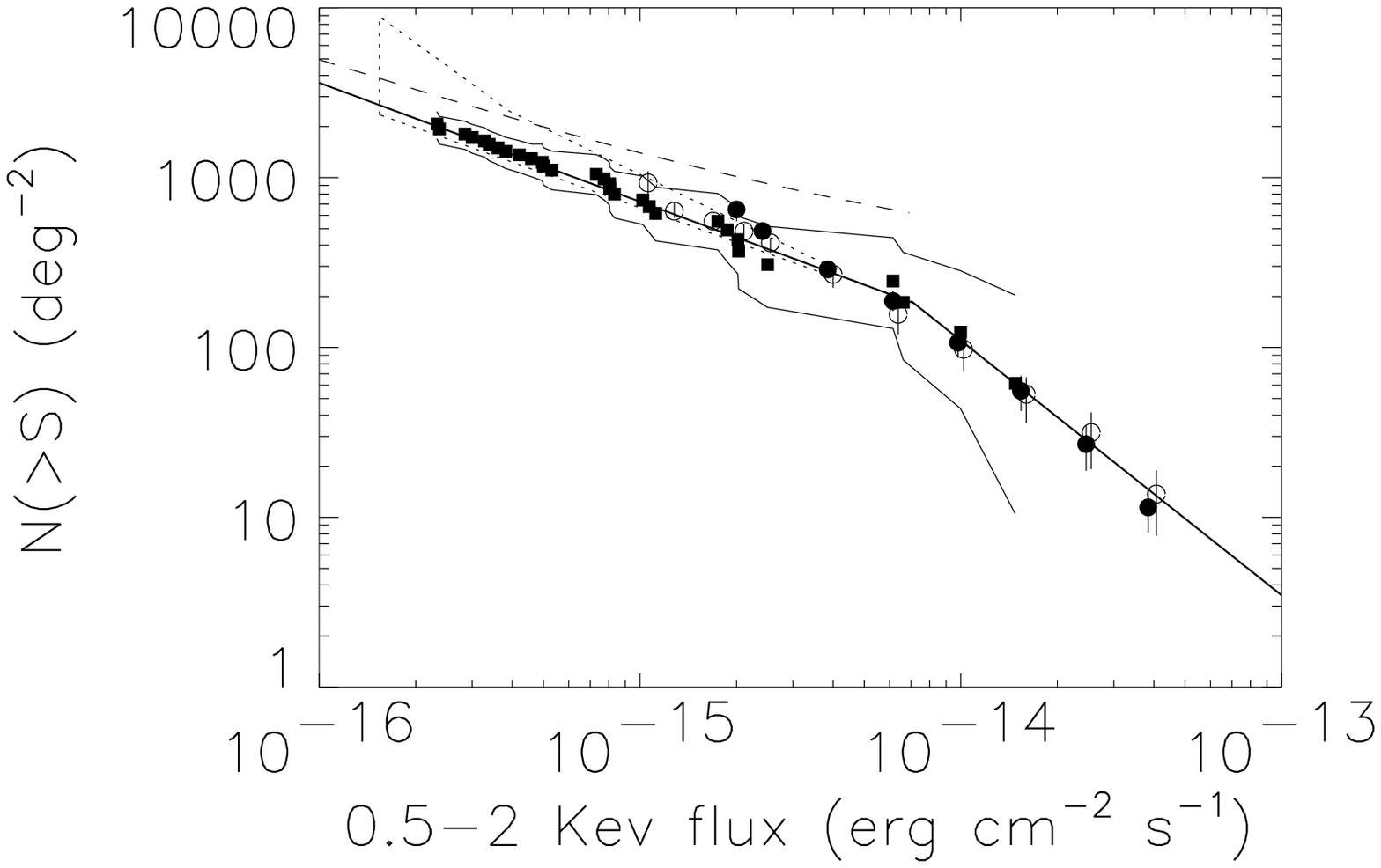,height=2.3in}}
\caption{
Integral number counts per square degree
of X-ray sources in the SSA13 field versus flux
for (a)\ the hard and (b)\ the soft energy bands.
The soft counts are based on 30 sources in the $10^{-7}$ probability sample
covering an area of 59 square arcminutes on the S3 chip.
The hard counts are based on the $10^{-7}$ probability sample of the S2 chip
in the $2-10$\ keV band and
a $10^{-7}$ probability sample of the S3 chip chosen in the $2-6$\ keV band
(to minimise background) and corrected to $2-10$\ keV fluxes.
The total area is 84 square arcminutes. At fluxes greater than
$2\times 10^{-14}$\ erg\ cm$^{-2}$\ s$^{-1}$, sources were drawn
from the S2, S3, I2, and I3 chips, which increased the area
to 227 square arcminutes. The combined hard sample contains 35 sources.
In (a)\ the solid line at bright fluxes is the
$N(>S)\propto S^{-3/2}$ representation of the
{\it ASCA} counts from Ueda et al.\cite{ueda2}
(sensitivity limit $7\times 10^{-14}$\ erg\ cm$^{-2}$\ s$^{-1}$);
these data lie on the extrapolation from previous results by
{\it HEAO1} A2\cite{picc} with a Euclidean slope of $-1.5$.
The dotted line shows the extrapolation of this line to fainter fluxes.
The solid line at fainter fluxes is the $-1.05$ power-law fit
to the present data below $2\times 10^{-14}$\ erg\ cm$^{-2}$\ s$^{-1}$.
The dashed line shows the normalisation at a given flux at which
integral counts with the observed shape would exceed a $2-10$\ keV sky
flux\cite{chen}
of $1.9\times 10^{-11}$\ erg\ cm$^{-2}$\ s$^{-1}$\ deg$^{-2}$.
In (b)\ the open and filled circles show the counts determined from
the {\it ROSAT} PSPC
(sensitivity limit $2\times 10^{-15}$\ erg\ cm$^{-2}$\ s$^{-1}$)
and HRI (sensitivity limit $10^{-15}$\ erg\ cm$^{-2}$\ s$^{-1}$)
data of the Lockman Hole from Hasinger et al.\cite{hasinger}.
The dotted line shows the fluctuation limits from
Hasinger et al.\cite{hasinger93}. The solid lines are the $-0.7$
index power-law fit to the data below
$7\times 10^{-14}$\ erg\ cm$^{-2}$\ s$^{-1}$
and the $-1.5$ index power-law fit above that flux.
The dashed line shows
the normalisation at a given flux at which integral counts with
the observed shape would exceed the $0.5-2$\ keV XRB\cite{chen}.
}
\label{figcounts}
\end{figure}

The cumulative counts per square degree, $N(>S)$,
are the sum of the inverse areas of all
sources brighter than flux $S$. Sources at the faintest
fluxes can be detected only at smaller off-axis angles
where the PSF and vignetting corrections are smaller; thus,
the area diminishes with flux.
In Fig.~\ref{figcounts}a, b we present our cumulative
counts per square degree (filled squares) in the soft and hard 
bands, respectively, with $1\sigma$
uncertainties from the Poisson error in the number of detected
sources (jagged solid lines). To the limiting flux levels
of $2.3\times 10^{-16}$\ erg\ cm$^{-2}$\ s$^{-1}$ ($0.5-2$\ keV) 
and $2.5\times 10^{-15}$\ erg\ cm$^{-2}$\ s$^{-1}$ ($2-10$\ keV),
simulations show that the counts are nearly complete and that
Eddington bias is unimportant; thus,
the raw counts accurately represent the true counts.

Our soft band counts are in excellent agreement with the deep
{\it ROSAT} counts in the Lockman Hole from
Hasinger et al.\cite{hasinger} in the region of overlap.
At fainter fluxes our new counts fall at the lower limit of their
fluctuation analysis, which suggests an ongoing flattening. 

An area-weighted maximum likelihood fit\cite{murdoch} of a single 
power-law to the 
$0.5-2$\ keV counts over the flux range 
$2.3-70\times 10^{-16}$\ erg\ cm$^{-2}$\ s$^{-1}$ is
given by the relation

\begin{equation}
N(>S)=185 \times (S/7 \times 10^{-15})^{-0.7 \pm 0.2}
\end{equation}

\noindent
where the errors on the power-law index are 68\% confidence.
Likewise, a power-law fit to the $2-10$\ keV counts over
the flux range $2.5-20\times 10^{-15}$\ erg\ cm$^{-2}$\ s$^{-1}$
is given by

\begin{equation}
N(>S)=170 \times (S/2 \times 10^{-14})^{-1.05 \pm 0.35}
\label{hardpleq}
\end{equation}

\noindent
where the counts intercept the {\it ASCA} extrapolation at the 
upper end of the flux range. Though the range
in indices is consistent with the power-law index of 1.5
seen at brighter fluxes, the counts are significantly lower 
than an extrapolation of the {\it ASCA} counts.

The source contributions to the XRB can be obtained by summing
the individual fluxes divided by area or, more indirectly, by
integrating $S dN$ using the power-law fits. We list the
directly summed source contributions to the XRB in the two bands 
in Table~2, along with previous determinations by {\it ROSAT}
and {\it ASCA}. With the additional 10 per cent contribution
from our data to the soft band, a maximum flux of
$1.1\times 10^{-12}$\ erg\ cm$^{-2}$\ s$^{-1}$\ deg$^{-2}$
remains to be accounted for. In the hard band, the combination
of the present results with the {\it ASCA} measurements at
higher fluxes means that at least $75$\ per cent of the background 
(using the highest published normalisation) is resolved to the 
currently observed flux limits.

\vskip 0.5cm
\centerline{\bf Optical Properties of the X-ray Sources}
\vskip 0.2cm

\begin{figure}[p]
\figurenum{2}
\centerline{\psfig{file=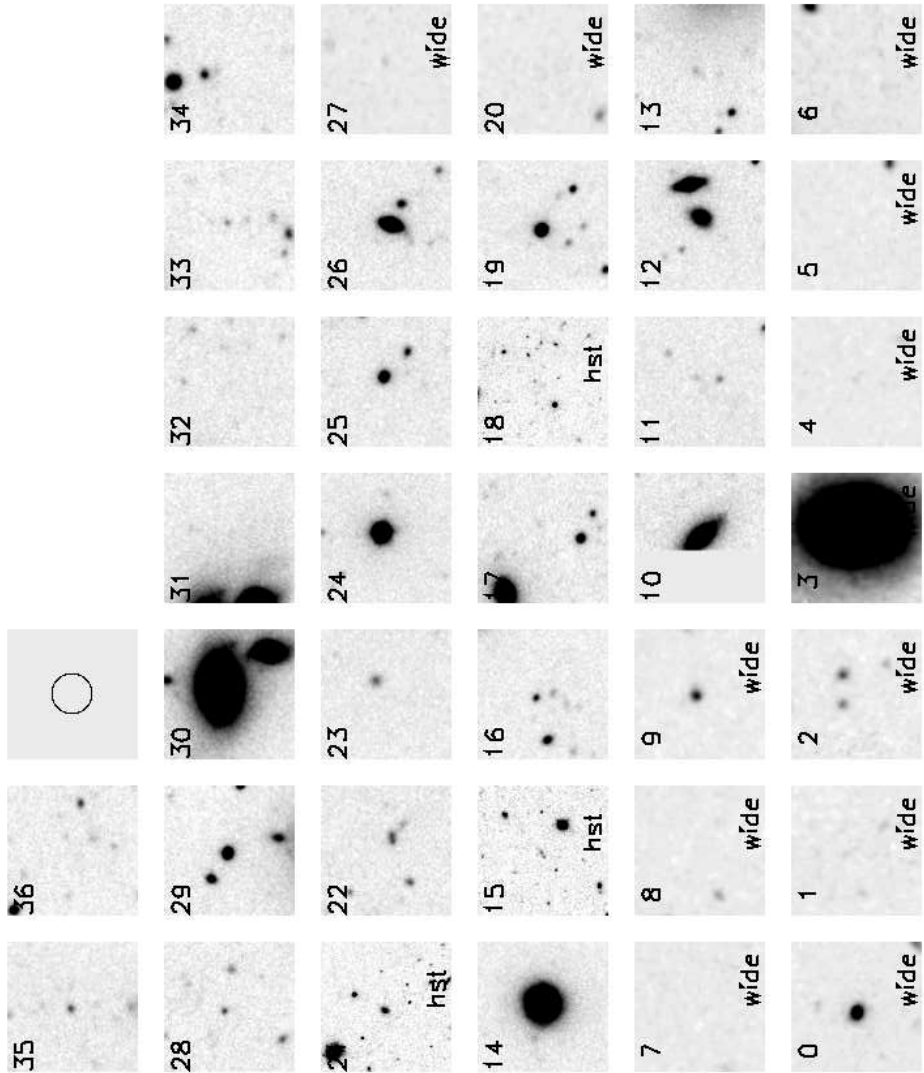,width=5.0in,angle=90}}
\caption{
$9''\times 9''$ $I$-band images of the X-ray sources of Table~1.
Most of the images are from ultradeep
data obtained with LRIS on the Keck 10\ m telescope, but those marked
wide are from shallower wide-field UH 2.2\ m data,
and those marked hst are from deep (approximately 16000 s of exposure)
F814W {\it HST} data. The astrometry of the optical images
is tied to deep 20\ cm VLA images currently being analysed by
Richards et al. (in preparation).
The absolute offset from the nominal {\it Chandra}
astrometry ($2.2''$\ W, $0.2''$\ N) was obtained from the quasar
CXO J131215.3+423901. A small adjustment to the pixel
size ($0.4908''$ versus $0.492''$) was also made to optimize the
agreement between the X-ray and optical sources, but no adjustment
of the roll angle. In this system the
{\it r.m.s.} dispersion between the 8 objects in the soft sample with
$I<23$ optical counterparts in the deep LRIS data is $0.36''$.
Intercomparison of the independent positions determined from the
hard and soft mages suggests that the error
may rise to as much as $1.7''$ in the faintest X-ray sources.
The ID numbers are as in Table~1, and the sources are ordered from
the lower left by right ascension.
The upper panel shows a circle of $1.5''$ radius
typical of the maximum positional uncertainty.
\label{figimages}
}
\end{figure}

We have compared our X-ray images with existing\cite{cowie96,wilson}
deep $HK'$, $I$, $B$, and $U'$
images obtained with the Keck 10\ m and UH 2.2\ m telescopes.
Because of the excellent $\sim 1$\ arcsec X-ray positional accuracy,
we can, in most cases, securely identify the optical counterparts
to the X-ray sources. In Fig.~\ref{figimages}a, b we show thumbnail
$I$-band images of all of the X-ray sources in Table~1.
Only one source,
CXO J131159.3+423928 (significant in both the hard and soft
X-ray images), is significantly extended in the X-ray images;
it is probably a high redshift cluster.
The optical image (thumbnail 35 of Fig.~\ref{figimages}a)
is centred on a faint ($I=23$) galaxy which lies at the centre
of a region of enhanced galaxy density.
In addition to the probable cluster, the hard sample contains
two quasars, eight bright galaxies, and eleven optically faint
($I>23$) objects, while the soft sample
contains five quasars, five bright galaxies, and fifteen 
unidentified optically faint objects.
Morphologically the X-ray selected bright galaxy population
consists of a mixture of early spirals and elliptical galaxies.
Three of the bright galaxies show possible signs of interaction
with nearby bright neighbors while the remainder are clearly
isolated.

\begin{figure}
\figurenum{3}
\vskip-2in
\centerline{\psfig{file=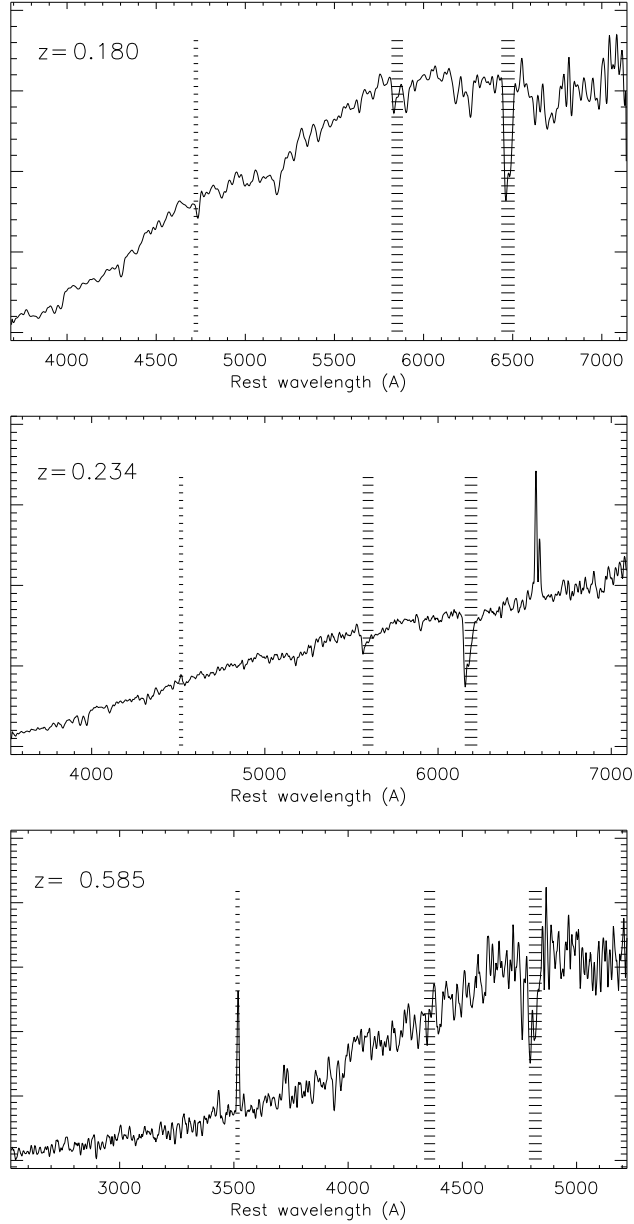,height=8.9in}}
\vskip-0.4in
\caption{
Keck LRIS spectra of three of the X-ray detected bright galaxies
that have redshifts (\#30 at $z=0.180$ and \#12 at $z=0.585$ are from
the hard-selected sample and \#14 at $z=0.234$ is from the
soft-selected sample). These objects do not show strong
emission features, except for H$\alpha$, NII, and SII in the
$z=0.234$ spectrum and [OII\}3727 in the $z=0.585$ spectrum.
The resolution of the spectra is $14$\ km/s and the shaded regions
show the positions of the strong 5577\ \AA\ night sky line and
the atmospheric bands.
\label{figspectra}
}
\end{figure}

Our Keck spectra for the quasars
show broad MgII or CIV and Lyman alpha lines.
In some of the bright galaxy spectra clear AGN signatures are 
present (e.g., a broad absorption line galaxy with 
P-Cygni profiles at $z=1.320$). However,
although subtle AGN signatures may be present in their optical
spectra, most of the bright galaxies
would not have been identified in an optical survey as AGN.
We illustrate this in Fig.~\ref{figspectra} with
spectra for three of the bright galaxies.

\vskip 0.5cm
\centerline{\bf The X-ray Spectrum}
\vskip 0.2cm

\begin{figure}[p]
\figurenum{4}
\centerline{\psfig{file=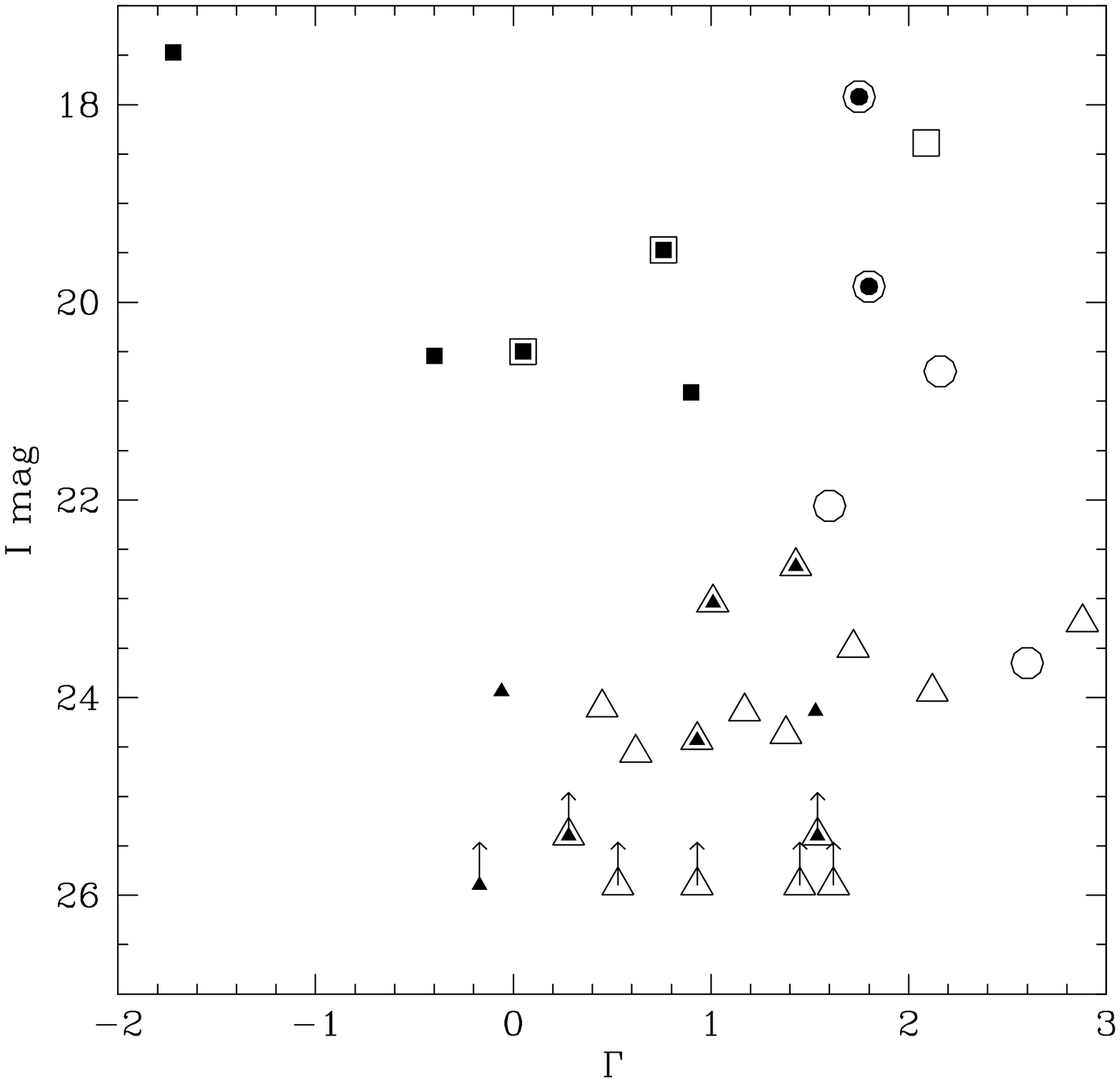,height=5.2in}}
\caption{
$I$ magnitudes versus X-ray photon indices.
Solid symbols represent the hard ($2-10$\ keV) selected sample,
and open symbols represent the soft ($0.5-2$\ keV) selected sample.
For objects with redshift identifications, quasars
are represented by circles and galaxies by squares.
Sources without redshifts are represented by triangles.
In cases where the objects were significantly detected in both samples,
the hard-selected magnitudes and indices were used; these objects
are indicated by solid symbols surrounded by larger open symbols.
\label{figmagvsindex}
}
\end{figure}

The photon intensity of the XRB, $P(E)$, where $E$ is the photon
energy in keV and $P(E)$ has units of
[photons\ cm$^{-2}$\ s$^{-1}$\ keV$^{-1}$\ sr$^{-1}$],
can be approximated by a power-law, $P(E)=AE^{-\Gamma}$.
The {\it HEAO1} A-2 experiment\cite{marshall} found that the XRB
spectrum from $3-15$\ keV was well described by a photon index
$\Gamma\simeq 1.4$, and this result has been confirmed and extended to
lower energies by recent analyses of {\it ASCA}\cite{gendreau,chen,
miyaji,ishisaki} and {\it BeppoSAX}\cite{vecchi} data.

The photon indices of the individual sources given in Table~1 were 
computed from the ratios of the counts in the $0.5-2$\ keV band
to those in the $2-10$\ keV band, assuming each source
could be described by a single power-law.
There is an extremely wide range of hardness in both samples,
ranging from negative indices to $\Gamma=1.8$ in the hard-selected
sample and from $\Gamma=0.1$ to values above 2 in the soft sample.
The composite (counts-weighted) photon index is $1.22\pm 0.03$
in the hard sample and $1.42\pm 0.04$
in the soft sample. The progressive hardening of the soft sample
as we move to fainter fluxes is a continuation of a trend seen
in the {\it ROSAT} samples\cite{hasinger93,almaini}.
The combined spectrum of all the soft X-ray sources of Table~1
is well fit by a single power-law over the $0.3-10$\ keV range
with an index of $1.42 \pm 0.07$ and an extinction corresponding
to the galactic $N(H)= 1.4 \times 10^{20}$ cm$^{-2}$.
If we assume that 75 per cent of the $2-10$\ keV
background has an index of 1.22 and that the remaining
25 per cent of the background comes from sources that have
fluxes greater than $1\times 10^{-13}$\ erg\ cm$^{-2}$\ s$^{-1}$
($2-10$\ keV) and an average photon index\cite{ueda1} of 1.63, 
then the index of the combined sample is 1.38, which agrees
extremely well with the spectrum of the hard X-ray background
\cite{marshall,gendreau,chen,miyaji,ishisaki,vecchi}.

Inspection of Table~1 suggests that the hardest sources
tend to correspond to the bright galaxies, with the optically
faint objects having intermediate hardness, and the quasars
being the softest of the sources observed.
We illustrate this more clearly in Fig.~\ref{figmagvsindex} where
we have plotted $I$-band magnitude versus photon index. Of the
$I\lesssim 22$ objects, more than half are galaxies,
and the majority of these have $\Gamma<1$. The five known quasars all
lie in the $\Gamma>1.7$ range, consistent with that of most
brighter AGN. The faint sources spread
over a wide range of indices that overlap both of the other populations.
We can quantify this by generating the counts-weighted averages
for each population separately. For the hard-selected sample,
we find that the bright galaxies (\#9, 12, 26, 29) have
an average photon index of $0.59 \pm 0.06$, the faint objects
(\#1, 6, 7, 8, 9, and 22) have $1.33 \pm 0.06$, and the
two quasars have $1.76 \pm 0.07$. For the soft-selected sample, 
the 15 unidentified objects with $I>23$ have a composite index of
$1.35 \pm 0.06$, which is almost identical to that of the
optically faint objects in the hard sample, and the quasars
have a composite index of $1.80 \pm 0.12$.

\vskip 0.5cm
\centerline{\bf The Source of the Background}
\vskip 0.2cm

Our data conclusively show that AGN are the major
contributors to the hard X-ray background.
Many of our sources agree with the predictions
of XRB synthesis
models\cite{setti,madau,matt,comastri,zdziarski,smith,gilli,schmidt2,miyaji2}
constructed within the framework of AGN
unification schemes to account for the spectral intensity
of the hard XRB and to explain
the X-ray source counts in the hard and soft energy bands.
In the unified scheme, the orientation of a molecular torus
surrounding the nucleus determines the classification of the
source. The models invoke, along with a population of unobscured
AGN, whose nuclear emission we see directly, a substantial
population of intrinsically obscured AGN whose hydrogen column
densities of $N_H\sim 10^{21}-10^{25}$\ cm$^{-2}$
around the nucleus block our line-of-sight.

The AGN that make up the hard XRB come in two main flavors:
roughly 40 per cent are luminous early-type galaxies (both
ellipticals and early spirals) in the
redshift range from $z=0$ to just beyond $z=1$, and
roughly 50 per cent have faint or, in some cases, undetectable
optical counterparts. Most of these objects would not have been
found even in sensitive optical surveys for AGN.

The bright galaxy population is extremely hard with an
average photon index of $\Gamma=0.59$. The X-ray sources
are point-like and centred on the galaxy nuclei, which suggests 
that they are produced by accretion onto the central black
holes that are known to be present in such systems.
The hardness of the X-ray spectra indicates that these X-ray
sources are highly obscured. Such sources were described by
Moran et al.\cite{moran} based on {\it Einstein} data. After 
hard X-ray components were discovered by Allen et al.\cite{allen} 
in {\it ASCA} spectra of six nearby giant elliptical 
galaxies, a model\cite{dm1} was constructed which
was able to account for a large fraction of the XRB 
with objects of this type. The model 
predicted that a significant fraction of the hard number 
counts at fluxes $<10^{-14}$\ erg\ cm$^{-2}$\ s$^{-1}$ could arise 
from sources at low redshift, as is indeed now observed to be the case.
The absolute $K$ magnitudes of these sources lie between
$\sim -24$ and $\sim -26$
(${\rm H_o}=65$\ km\ s$^{-1}$\ Mpc$^{-1}$ and $q_o=0.5$),
or from just below to several times the $L_{\ast}$ luminosity,
and their rest-frame $2-10$\ keV luminosities range from $5\times
10^{41}$ to $3\times 10^{43}$\ erg\ s$^{-1}$.
These sources are at too low redshifts to be likely
submillimeter candidates; however, they should be far-infrared
sources, which SIRTF and other upcoming airborne
and space missions should be able to detect.

The optically faint sources have an average photon
index of $\Gamma=1.3$. These sources could either be a
smooth continuation to $z>1$ of the bright early-type
galaxies with obscured luminous X-ray nuclei, more
distant obscured AGN, or something more exotic,
such as extremely high redshift ($z\gg 5$) quasars.
For this final possibility, the objects would be invisible in the 
$B$-band because of scattering by the foreground intergalactic 
neutral hydrogen.
In the soft sample, eleven of the optically faint sources have
$B>26$ and could lie in this category. This places an upper
limit on the surface density of this type of source of 0.26 per
square arcminute to the $3\times10^{-16}$\ erg\ cm$^{-2}$\ s$^{-1}$
limit of the $0.5-2$\ keV sample, which is slightly lower than
the predictions of the toy model of Haiman and Loeb\cite{haiman}
for X-ray selected high-redshift quasars.
The handful of objects which are detected in the NIR
but absent in $B$ are the most promising candidates for this
type of object and, in some cases (e.g., object 19 in the soft
sample) may be bright enough for follow-up with NIR
spectroscopy to test the hypothesis.

Comparison with submillimeter\cite{barger1,hughes,gunn}, 
far-infrared, and radio
samples should allow us to determine what fraction of
objects in these surveys are X-ray emitting AGN.
As near-infrared spectra and photometric redshift estimates of 
the optically faint sources are established, 
we will be able to refine the obscured AGN models and determine
whether any of the faint sources are indeed very high redshift
quasars. With the X-ray, 
optical, and submillimeter samples all now approaching the full 
resolution of their respective backgrounds, we are close to achieving 
a complete cosmic census of the population of galaxies and AGN.

\vskip 0.5cm
\centerline{\bf Acknowledgements}
\vskip 0.2cm

     We thank J. Halpern and G. Hasinger for comments which
     greatly improved the first draft of this paper.
     We acknowledge E. Boldt for his many years of pioneering
     work concerning the X-ray background and 
     R. Giacconi, whose insight and enthusiasm have
     inspired this subject. We thank the CXC, L. Van Speybroeck, 
     M. Weisskopf and the MSFC team, M. Bautz, G. Garmire, and 
     the ACIS team for building and operating such an excellent 
     observatory. We acknowledge the use of HEASARC software.
     A.J. Barger acknowledges support from
     the Hubble and Chandra fellowship programs.

\newpage

\noindent
{\bf Notes to Table~1}
X-ray sources in the SSA13 field selected in the hard 
($2-10$\ keV; S2 and S3 chips)
or soft ($0.5-2$\ keV; S3 chip) bands.
The X-ray images were prepared using xselect 
and associated ftools at GSFC. ACIS grades 0, 2, 3, 4, and 6 were 
used, and columns at the boundaries of the readout nodes were rejected.
Counts lying within a $5''$ diameter aperture
were measured, together with the background in a
$5''-7.5''$ radius annulus, at $2''$ intervals along the field.
The distribution of detected counts is Poisson.
A cut of 17 counts in the hard S3 image
and 10 counts in the other two images represents a 
$<10^{-7}$ probability threshold against background fluctuations and 
ensures a $<20$\% probability of a single spurious
source detection in the entire sample. 
The source counts were corrected for the enclosed energy fraction
within the aperture. For the S3 chip the flux calibrations
were made using an array of effective areas versus energy at 12
positions and an assumed power-law spectrum having counts-weighted
mean photon indices $\Gamma=1.2$ ($2-10$\ keV) and
$\Gamma=1.4$ ($0.5-2$\ keV).
The galactic $N(H)= 1.4 \times 10^{20}$ cm$^{-2}$ is too
low to affect the flux conversions.
For the S2 chip a single conversion factor of
$2.6\times 10^{-11}$\ erg\ cm$^{-2}$\ ct$^{-1}$ was used.
Using the on-axis flux calibrations of
$2.5\times 10^{-11}$ ($2-10$\ keV)
and $2.9\times 10^{-12}$ ($0.5-2$\ keV)
to convert the S3 counts per second to flux,
we determine limiting minimum fluxes of
$3.2\times 10^{-15}$\ erg\ cm$^{-2}$\ s$^{-1}$ ($2-10$\ keV)
and $3.0\times 10^{-16}$\ erg\ cm$^{-2}$\ s$^{-1}$ ($0.5-2$\ keV).
The table is restricted to sources with off-axis angles 
$<6'$, where $>50$\% of the energy is enclosed within a $\sim2.5''$ 
radius, and to sources where the noise, computed from the variance of
the background and signal, is less than one third the signal.
The $15''$ borders of each chip have incomplete
exposure times due to the spacecraft dither so objects detected
in these borders were not included in our counts analysis.
The NIR and optical magnitudes are computed in $1.5''$
radii intervals. $I$ is Kron-Cousins, $B$ is Johnson,
$HK'$ is a broad filter centred at $1.9$
microns, and $U'$ is a $300$\ \AA\
filter centred at $3400$\ \AA. Lower limits are $1\sigma$.

\noindent
{\bf Notes to Table~2}
The statistical errors on our observed sky brightnesses dominate
the systematic errors, which are expected to be less than 
10 per cent. To be consistent with Hasinger et al.\cite{hasinger},
we converted our soft band sky brightness of
$6.0\pm 1.5\times 10^{-13}$\ erg\ cm$^{-2}$\ s$^{-1}$\ deg$^{-2}$
to the $1-2$\ keV range using the measured mean photon index.
This result was then compared with the $1-2$\ keV background
(where galactic contamination is less than at lower energies) of
$3.7-4.4\times 10^{-12}$\ erg\ cm$^{-2}$\ s$^{-1}$\ deg$^{-2}$
from Gendreau et al.\cite{gendreau}, using a fit to {\it ASCA}
data, and Chen et al.\cite{chen}, using a fit to joint
{\it ASCA}/{\it ROSAT} data.
In the hard band the summed counts are compared with
the $2-10$\ keV background of
$1.6-2.3\times 10^{-11}$\ erg\ cm$^{-2}$\ s$^{-1}$\ deg$^{-2}$
from Marshall et al.\cite{marshall},
using a fit to {\it HEAO1} A2 data, and Vecchi et al.\cite{vecchi},
using a fit to {\it BeppoSAX} data.

\newpage

\begin{deluxetable}{lrrrrrrrrrrrrrr}
\tablewidth{0pt}
\scriptsize
\tablenum{1}
\tablehead{
\colhead{\#} & \multicolumn{3}{c}{RA(2000)} &
\multicolumn{3}{c}{Dec(2000)} &
$f$(2--10\ keV) & $f$(0.5--2\ keV) & \colhead{$\Gamma$} &
\colhead{$HK'$} & \colhead{$I$} & \colhead{$B$} & \colhead{$U'$} &
\colhead{$z$} \cr
& & & & & & & ($10^{-16}$\ cgs) & ($10^{-17}$\ cgs) & & & & & &
}
\startdata
0\tablenotemark{S2} & 13 & 12 & 43.38 & 42 & 44 & 36.73 & $82.5 \pm 21.0$ & \nodata
& \nodata & 20.13 & 24.60 & \nodata & \nodata & \nodata \cr
1\tablenotemark{b} & 13 & 12 & 40.24 & 42 & 39 & 35.55 & $43.0 \pm 14.1$ & $79.0 \pm 16.7$ & 0.93
& \nodata & 24.43 & 25.19 & \nodata & \nodata \cr
2\tablenotemark{S2} & 13 & 12 & 39.62 & 42 & 45 & 48.77 & $89.9 \pm 25.2$ & \nodata
& \nodata & 24.36 & $>26.7$ & \nodata & \nodata & \nodata \cr
3\tablenotemark{S2} & 13 & 12 & 39.50 & 42 & 42 & 48.83 & $39.2 \pm 11.4$ & \nodata
& \nodata & 17.08 & 19.89 & \nodata & \nodata & 0.111 \cr
4\tablenotemark{s} & 13 & 12 & 37.94 & 42 & 40 & 5.53 & $32.1\pm 11.8$ & $32.8\pm 10.8$ & 0.45
& \nodata & 24.10 & $25.76$ & \nodata & \nodata \cr
5\tablenotemark{S2} & 13 & 12 & 37.16 & 42 & 43 & 21.08 & $50.0 \pm 12.5$ & \nodata
& \nodata & $>25.4$ & $>26.7$ & \nodata & \nodata & \nodata\cr
6\tablenotemark{b} & 13 & 12 & 36.86 & 42 & 38 & 44.55 & $46.2 \pm 14.2$ & $39.2 \pm 12.1$ & 0.28
& \nodata & $>25.4$ & $>26.7$ & \nodata & \nodata \cr
7\tablenotemark{b} & 13 & 12 & 36.58 & 42 & 40 & 2.80 & $384 \pm 34.0$ & $1480 \pm 67.1$ & 1.54
& \nodata & $>25.4$ & $26.70$ & \nodata & \nodata \cr
8\tablenotemark{h} & 13 & 12 & 36.00 & 42 & 40 & 44.11 & $41.6 \pm 12.3$ & $23.1 \pm 9.22$ & $-0.06$
& \nodata & 23.94 & 25.40 & $>25.5$ & \nodata \cr
9\tablenotemark{b,S3e} & 13 & 12 & 35.68 & 42 & 41 & 50.67 & $232 \pm 26.0$ & $408 \pm 35.3$ & 0.90
& \nodata & 20.91 & 22.95 & 23.40 & 1.320 \cr
10\tablenotemark{S2} & 13 & 12 & 34.48 & 42 & 43 & 9.27 & $78.1 \pm 14.9$ & \nodata
& \nodata & 16.38 & 19.26 & 23.22 & 24.19 & 0.241 \cr
11\tablenotemark{s} & 13 & 12 & 32.36 & 42 & 39 & 49.39 & $15.7\pm 8.86$ & $50.0\pm 12.9$ & 1.38
& 20.16 & 24.37 & 25.82 & $>25.5$ & \nodata \cr
12\tablenotemark{h} & 13 & 12 & 31.34 & 42 & 39 & 2.19 & $48.4 \pm 13.1$ & $19.8 \pm 8.60$ & $-0.40$
& 18.02 & 20.54 & 23.37 & 23.76 & 0.586 \cr
13\tablenotemark{s} & 13 & 12 & 30.83 & 42 & 39 & 42.73 & $4.71\pm 6.69$ & $38.2\pm 11.3$ & 2.12
& 19.75 & 23.94 & 25.80 & 25.36 & \nodata \cr
14\tablenotemark{s} & 13 & 12 & 29.26 & 42 & 37 & 32.33 & $14.5\pm 10.6$ & $112\pm 20.1$ & 2.09
& 16.04 & 18.39 & 20.91 & 21.35 & 0.234 \cr
15\tablenotemark{S2} & 13 & 12 & 28.25 & 42 & 44 & 54.52 & $56.7 \pm 13.7$ & \nodata
& \nodata & 21.64 & 24.41 & 25.33 & $>25.5$ & \nodata \cr
16\tablenotemark{b} & 13 & 12 & 26.00 & 42 & 37 & 35.86 & $49.9 \pm 15.2$ & $170 \pm 24.6$ & 1.43
& 19.73 & 22.67 & 24.53 & 25.14 & \nodata \cr
17\tablenotemark{s} & 13 & 12 & 25.29 & 42 & 41 & 19.53 & $13.7\pm 8.14$ & $49.6\pm 12.8$ & 1.45
& 21.68 & $>25.9$ & 26.93 & $>25.5$ & \nodata \cr
18\tablenotemark{S2e} & 13 & 12 & 22.48 & 42 & 44 & 49.97 & $42.1 \pm 11.7$ & \nodata & \nodata
& 20.73 & 24.98 & 26.26 & $>25.5$ & \nodata \cr
19\tablenotemark{b} & 13 & 12 & 22.32 & 42 & 38 & 13.89 & $116 \pm 19.8$ & $622 \pm 44.6$ & 1.80
& 17.80 & 19.84 & 21.53 & 22.23 & 2.565\tablenotemark{q} \cr
20\tablenotemark{s} & 13 & 12 & 21.63 & 42 & 35 & 49.97 & $45.4\pm 28.4$ & $203\pm 33.9$ & 1.65
& \nodata & 25.06 & $>26.7$ & \nodata & \nodata \cr
21\tablenotemark{s} & 13 & 12 & 21.50 & 42 & 44 & 5.41 & $17.3\pm 8.92$ & $78.7\pm 16.1$ & 1.60
& 19.23 & 22.06 & 23.36 & 23.98 & 1.305\tablenotemark{q} \cr
22\tablenotemark{b} & 13 & 12 & 20.11 & 42 & 42 & 22.42 & $49.0 \pm 12.3$ & $196 \pm 24.6$ & 1.53
& $>22.5$ & 24.14 & 25.75 & $>25.5$ & \nodata \cr
23\tablenotemark{s} & 13 & 12 & 19.19 & 42 & 38 & 8.36 & $28.3\pm 11.5$ & $36.1\pm 11.6$ & 0.62
& $>22.5$ & 24.56 & 26.72 & $>25.5$ & \nodata \cr
24\tablenotemark{b} & 13 & 12 & 15.32 & 42 & 39 & 0.22 & $190 \pm 23.3$ & $986 \pm 54.4$ & 1.75
& 16.18 & 17.92 & 18.66 & 19.00 & 2.565\tablenotemark{q} \cr
25\tablenotemark{s} & 13 & 12 & 11.72 & 42 & 44 & 12.59 & $19.5\pm 10.0$ & $175\pm 24.3$ & 2.16
& 18.65 & 20.70 & 22.22 & 22.27 & 0.950\tablenotemark{q} \cr
26\tablenotemark{b} & 13 & 12 & 10.02 & 42 & 41 & 29.94 & $151 \pm 20.3$ & $246 \pm 27.8$ & 0.76
& 16.74 & 19.47 & 22.50 & 24.15 & 0.212 \cr
27\tablenotemark{s} & 13 & 12 & 9.93 & 42 & 36 & 15.30 & $15.9\pm 25.8$ & $77.3\pm 22.2$ & 1.72
& 19.66 & 23.50 & 27.32 & $>25.5$ & \nodata \cr
28\tablenotemark{s} & 13 & 12 & 8.38 & 42 & 41 & 43.08 & $2.37\pm 6.18$ & $53.1\pm 13.4$ & 2.88
& 20.39 & 23.24 & 25.64 & $>25.5$ & \nodata \cr
29\tablenotemark{b} & 13 & 12 & 6.55 & 42 & 41 & 41.31 & $138 \pm 20.2$ & $95.8 \pm 17.7$ & 0.05
& 17.80 & 20.50 & 23.66 & 24.82 & 0.696 \cr
30\tablenotemark{h} & 13 & 12 & 6.55 & 42 & 41 & 25.16 & $38.4 \pm 11.8$ & $8.17 \pm 6.45$ & $-1.72$
& 15.22 & 17.47 & 20.52 & 22.31 & 0.180 \cr
31\tablenotemark{s} & 13 & 12 & 5.18 & 42 & 41 & 23.45 & $7.29\pm 7.77$ & $33.8\pm 11.3$ & 1.62
& 21.16 & $>25.9$ & 28.02 & $>25.5$ & \nodata \cr
32\tablenotemark{h} & 13 & 12 & 4.18 & 42 & 41 & 13.59 & $48.2 \pm 13.7$ & $25.9 \pm 10.2$ & $-0.17$
& 22.39 & $>25.9$ & $>27.6$ & $>25.5$ & \nodata \cr
33\tablenotemark{s} & 13 & 12 & 1.23 & 42 & 42 & 7.78 & $5.28\pm 8.64$ & $83.3\pm 17.9$ & 2.60
& 21.08 & 23.65 & 26.67 & $>25.5$ & 3.405\tablenotemark{q} \cr
34\tablenotemark{s} & 13 & 11 & 59.66 & 42 & 41 & 52.89 & $34.3\pm 13.9$ & $42.1\pm 13.5$ & 0.53
& \nodata & $>25.9$ & 26.21 & $>25.5$ & \nodata \cr
35\tablenotemark{b} & 13 & 11 & 59.36 & 42 & 39 & 28.17 & $273 \pm 36.5$ & $612 \pm 51.2$ & 1.01
& \nodata & 23.04 & $>27.6$ & $>25.5$ & \nodata \cr
36\tablenotemark{s} & 13 & 11 & 59.19 & 42 & 38 & 34.12 & $55.5\pm 22.6$ & $107\pm 23.3$ & 0.93
& \nodata & $>25.9$ & 27.29 & $>25.5$ & \nodata \cr
\enddata
\tablenotetext{s}{S3 source detected only in soft band sample.}
\tablenotetext{h}{S3 source detected only in hard band sample.}
\tablenotetext{b}{Significant detection in both samples.}
\tablenotetext{S3e}{In the S3 $15''$ excluded border.}
\tablenotetext{S2}{Detected in the S2 chip.}
\tablenotetext{S2e}{In the S2 $15''$ excluded border.}
\tablenotetext{q}{Quasar spectrum.}
\end{deluxetable}

\clearpage

\begin{deluxetable}{ccccc}
\tablewidth{0pt}
\tablenum{2}
\tablecaption{Source Contributions to the XRB\label{tab2}}
\tablehead{
\colhead{Energy range} &
\colhead{Flux range} & \colhead{Source Contribution} &
\colhead{Percentage} & \colhead{Reference} \\
\colhead{(keV)} & \colhead{(erg\ cm$^{-2}$\ s$^{-1}$)} &
\colhead{(erg\ cm$^{-2}$\ s$^{-1}$\ deg$^{-2}$)} & 
\colhead{of XRB\tablenotemark{1}} &
}
\startdata
$1-2$ & $>10^{-15}$ & $3.0\times 10^{-12}$ & $68-81$ &
Hasinger et al.\cite{hasinger} \\
$1-2$ & $(2.3-10)\times 10^{-16}$ & $(3.8\pm 1.0)\times 10^{-13}$ & $6-13$ &
present paper\\
$2-10$ & $>10^{-13}$ & $4.5\times 10^{-12}$ & $20-28$ &
Ueda et al.\cite{ueda1}\\
$2-10$ & $(2.5-100)\times 10^{-15}$ & $(1.30\pm 0.3)\times 10^{-11}$ &
$56-81$ & present paper\\
\enddata
\tablenotetext{1}{The range given is a combination of the uncertainty
in the source contribution (from the third column) and the variation in
the published sky flux.}
\end{deluxetable}

\end{document}